\documentclass[dvips]{elsart}

\usepackage[dvips]{graphicx}
\graphicspath{{img/}}
\usepackage{amssymb}
\usepackage[fleqn]{amsmath}

\newcommand{\abs}[1]{\left|#1\right|}
\renewcommand{\vec}[1]{\boldsymbol{#1}}
\newcommand{\vev}[1]{\langle#1\rangle}
\def\bra#1{\langle#1\rvert}
\def\ket#1{\lvert#1\rangle}
\DeclareMathOperator{\sgn}{sgn}
\DeclareMathOperator{\Tr}{Tr}

\def\t{\Tilde}
\def\wt{\widetilde}
\def\dt#1{\wt{\wt{#1}}}
\def\p{{\vec p}}
\def\q{{\vec q}}
\def\qc{{\q_c}}
\def\clean{{\text{clean}}}
\def\cutoff{{\text{cutoff}}}
\def\ladder{{\text{ladder}}}
\def\cross{{\text{crossed}}}
\def\dirty{{\text{dirty}}}
\def\sp{{\text{sp}}} 
\renewcommand{\O}{\mathcal{O}}

\newcommand{\fig}[4][clip]{%
  \begin{figure}[htbp]%
    \centering\includegraphics*[#1]{#2}%
    \caption{#4}\label{#3}%
  \end{figure}}

\newcommand{\figs}[5][clip]{%
  \begin{figure}[htbp]%
    \centering\textbf{(a)} \includegraphics*[#1]{#2}%
    \hspace{1cm}\textbf{(b)} \includegraphics*[#1]{#3}%
    \caption{#5}\label{#4}%
  \end{figure}}

\begin{document}
\begin{frontmatter}
\title{Disordered loops in the two-dimensional \\
  antiferromagnetic spin-fermion model}
\author{T. Enss, S. Caprara, C. Castellani, C. Di Castro, and M. Grilli}
\address{CNR--INFM--SMC Center and 
  Dipt.\ di Fisica, Univ.\ di Roma ``La Sapienza'', \\
  P.le A. Moro 5, 00185 Roma, Italy}

\begin{abstract}
  The spin-fermion model has long been used to describe the
  quantum-critical behavior of 2d electron systems near an
  antiferromagnetic (AFM) instability.  Recently, the standard
  procedure to integrate out the fermions to obtain an effective
  action for spin waves has been questioned in the clean case.  We
  show that in the presence of disorder, the single fermion loops
  display two crossover scales: upon lowering the energy, the
  singularities of the clean fermionic loops are first cut off, but
  below a second scale new singularities arise that lead again to
  marginal scaling.  In addition, impurity lines between different
  fermion loops generate new relevant couplings which dominate at low
  energies.  We outline a non-linear $\sigma$ model formulation of the
  single-loop problem, which allows to control the higher
  singularities and provides an effective model in terms of low-energy
  diffusive as well as spin modes.
\end{abstract}

\end{frontmatter}


\section{Introduction}
\label{sec:intro}

The spin-fermion model is a low-energy effective model describing the
interaction of conductance electrons (fermions) with spin waves
(bosons).  It has been used, e.g., to describe the quantum critical
behavior of an electron system near an antiferromagnetic instability
\cite{Her76,Mil93,Sac99}.  An important example where this might be
realized experimentally is in itinerant heavy-fermion materials
\cite{Ste01}.  By integrating out the fermions completely, a purely
bosonic effective action for spin waves is obtained.  This action is
written in terms of a bare spin propagator and bare bosonic vertices
with any even number of spin lines.  The value of each bosonic vertex
is given by a fermionic loop with spin-vertex insertions: in general,
these are complicated functions of all external bosonic frequencies
and momenta.  Hertz \cite{Her76} and Millis \cite{Mil93} considered
only the static limit of these vertices, i.e., setting all frequencies
to zero at finite momenta.  In this limit the 4-point vertex and all
higher vertices vanish for a linear dispersion relation, while they
are constants proportional to a power of the inverse bandwidth if the
band curvature is taken into account.  For the AFM the dynamic
critical exponent is $z=2$ due to Landau damping of spin modes by
particle-hole pairs.  The scaling of these vertices in $d=2$ under an
RG flow toward low energy scales is marginal for the 4-point vertex
while all higher vertices are irrelevant ($d+z=4$ is the
upper-critical dimension).  Thus, a well-defined bosonic action with
only quadratic and quartic parts in the spin field is obtained.

Recently, Lercher and Wheatley \cite{LW00} as well as Abanov
\textit{et al.} \cite{ACS03} considered for the 2d case not only the
static limit of the 4-point vertex but the full functional dependence
on frequencies and momenta.  Surprisingly, they found that in the
\emph{dynamic limit}, setting the momenta to zero at finite frequency,
the 4-point vertex is strongly divergent as the external frequencies
tend to zero, implying an effective spin interaction nonlocal in time.
The higher bosonic vertices display an even stronger singularity
\cite{AC04}.

To assess the relevance of the singular bosonic vertices,
Ref.~\cite{AC04} considered the \emph{scaling limit} $\omega \sim q^2$
with $z=2$.  In this limit, the bosonic vertices are less singular
than in the dynamic limit but the related couplings are still
marginal, i.e., they cannot be neglected in the effective bosonic
action, in apparent contradiction to Hertz and Millis.  Employing an
expansion in a large number of hot spots $N$ or fermion flavors,
Ref.~\cite{ACS03} argues that vertex corrections are resummed to yield
a spin propagator with an anomalous dimension $\eta = 2/N = 1/4$ (for
$N=8$).  At the same time, $z=2$ remains unchanged up to two-loop
order, i.e., the frequency dimension is given by $x_\omega =
2(1-\eta/2)$.

The infinite number of marginal vertices renders the purely bosonic
theory difficult to use for perturbative calculations.  A relevant
question is whether the above difficulty persists upon the inclusion
of a weak static disorder potential present in real materials.  To
address this issue we insert disorder corrections into single
fermionic loops and find two different crossover scales: at
frequencies $\omega \gg 1/\tau$ (i.e., much larger than the impurity
scattering rate) and momenta $q \gg 1/\ell$ (with mean free path
$\ell=v_F\tau$, where $v_F$ is the Fermi velocity), the fermionic
loops resemble the clean case, while below this scale the singularity
is cut off by self-energy corrections and the loops saturate.
However, at yet lower frequencies, a second crossover scale $\omega
\sim 1/(\tau k_F\ell)$, $q \sim 1/(\ell\sqrt{k_F\ell})$ appears where
the loops acquire a diffusive form due to impurity ladder corrections
and the related couplings again scale marginally, as in the clean
case.  Therefore, in an intermediate energy range disorder regularizes
the singular vertices and appears to restore Hertz and Millis theory,
while ultimately at the lowest scales the disordered loops are as
singular as the clean ones, albeit with a different functional form:
the linear dispersion of the electrons is replaced by a diffusive
form.  We outline a non-linear $\sigma$ model formulation of the
disordered single-loop problem which allows us to identify all
disorder corrections which exhibit the maximum singularity, and
provides an action for spin modes coupled to low-energy diffusive
electronic modes, instead of the original electrons.

Finally, while all disorder corrections to a single fermion loop lead
to couplings which scale at most marginally, impurity lines connecting
different fermion loops are a relevant perturbation in $d=2$.  We find
that these diagrams may dominate the single-loop contributions below
$\omega\simeq 1/\tau$, depending on the typical values of the bosonic
momenta.

We proceed as follows: in the remaining part of this section, we
introduce the model and the scaling arguments for the clean case.  We
then insert disorder corrections into a single fermion loop and
discuss a class of most singular diagrams in section \ref{sec:disord}.
Their scaling behavior and the emergence of two crossover scales is
the subject of section \ref{sec:scale}.  The multi-loop diagrams are
discussed in section \ref{sec:multi}.  Appendix \ref{sec:nlsm}
contains the non-linear $\sigma$ model for the disordered single-loop
case.

\subsection{The spin-fermion model in 2d}

The 2d spin-fermion model is defined by the action
\begin{align*}
  S[\psi,\bar\psi,\phi] & = (\bar\psi,G_0^{-1}\psi) +
  (\phi,\chi_0^{-1}\phi) + g\phi\bar\psi\psi
\end{align*}
for a fermionic field $\psi$, $\bar\psi$ and a bosonic spin field
$\phi$.\footnote{The explicit spin structure of the spin-fermion
  vertex is not relevant for the fermionic loops and needs to be
  specified only when spin lines are contracted.}  The inverse
fermionic propagator is $G_0^{-1}(i\epsilon,\p) = i\epsilon-\xi_\p$ in
terms of the Matsubara frequency $i\epsilon$ and a dispersion relation
$\xi_\p$ with a roughly circular Fermi surface (FS), which we
approximate by a quadratic dispersion $\xi_\p = \frac{\abs\p^2} {2m_e}
- \mu$ with electron mass $m_e$, chemical potential $\mu$, Fermi
momentum $k_F = \sqrt{2m_e\mu}$, and constant density of states $2\pi
\rho_0 = \epsilon_F/v_F^2$.  $\chi_0(\q)$ is the bare spin propagator.

We assume that the above model describes an AFM quantum critical point
at finite $\q=\qc$.  The Fermi surface has so-called hot regions
connected by exchange of $\qc$, and cold regions where scattering off
spin waves is weak.  Here we shall assume an underlying lattice and a
commensurate $\qc=(\pi,\pi)$, which is equivalent (up to a reciprocal
lattice vector) to $-\qc$.

\fig[height=6cm,clip]{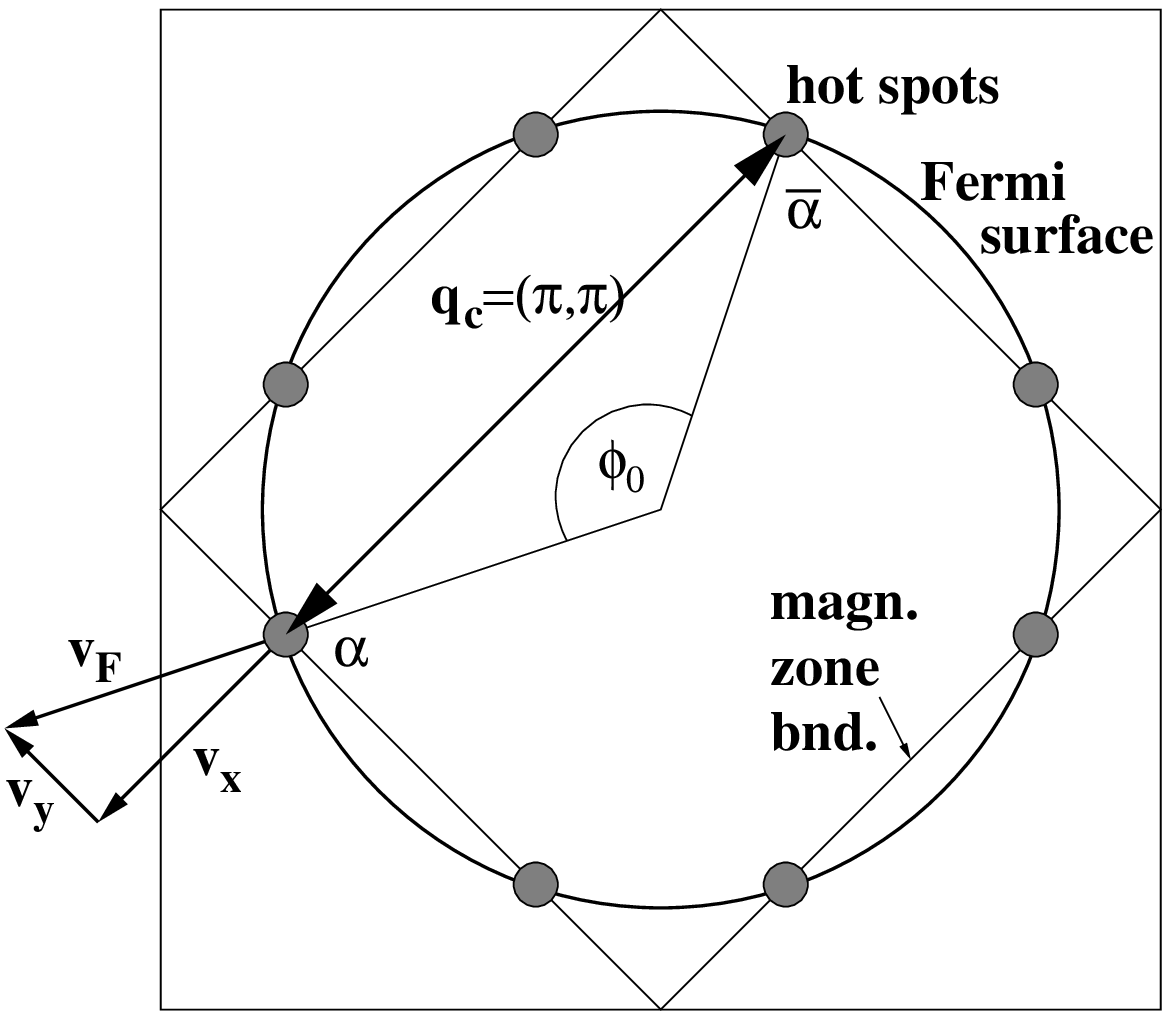}{fig:FS}{The Fermi surface with the hot
  spots separated by the wave vector $\qc$.}

When computing fermionic loops with only spin-vertex insertions, the
momentum integration can be reduced to the region around two hot spots
separated by $\qc$, which we shall denote by $\alpha$ and $\bar\alpha$
(Fig.~\ref{fig:FS}).  The fermionic dispersion relation $\xi_\p$ is
linearized around any hot spot $\alpha$ at momentum
$\p_{\text{hs}}^\alpha$ as \cite{ACS03}
\begin{align*}
  \xi_\p & = \vec{v}_F (\p-\p_{\text{hs}}^\alpha) 
  = v_x^\alpha \bar p_x + v_y^\alpha \bar p_y 
  \equiv \xi_{\bar\p}^\alpha
\end{align*}
where $\bar\p$ denotes the distance from the hot spot.  The components
$v_x$ ($v_y$) of the Fermi velocity $v_F$ parallel (perpendicular) to
$\qc$ at a given hot spot $\alpha$ are related by $v_F^2 = v_x^2 +
v_y^2$ and $v_x/v_y = \tan(\phi_0/2)$, with $\phi_0$ the angle between
hot spots $\alpha$ and $\bar\alpha$ as seen from the center of the
circular Fermi surface.  The case $\phi_0=\pi$ ($v_y=0$) corresponds
to perfect nesting, but here we consider a generic $\phi_0$ without
nesting.  For a pair of hot spots, the momentum integration can be
written as
\begin{align*}
  \int \frac{d^2p}{(2\pi)^2}
  & = J \sum_\alpha \int d\xi^\alpha \, d\xi^{\bar\alpha}
\end{align*}
where $\xi^\alpha$ and $\xi^{\bar\alpha}$ are two independent momentum
directions at hot spot $\alpha$ ($\xi^{\bar\alpha}$ coincides with the
radial direction at hot spot $\bar\alpha$), $J^{-1} = 4\pi^2 v_F^2
\sin\phi_0$ is the corresponding Jacobian which depends on the shape
of the Fermi surface and the filling, and one still has to perform the
summation over all $N=8$ hot spots.

\subsection{Clean fermionic $2n$-loops}

The fermionic loops with $2n$ spin-vertex insertions---i.e., the
$2n$-point functions---are in general complicated functions of the
external frequencies and momenta.  The loop with two spin insertions
contributes to the self-energy for the spin propagator and has the
well-known Landau damping form for small frequencies $\omega$ and
momenta near $\qc$ \cite{Ste67},
\begin{align*}
  \Sigma(i\omega,\q\approx\qc) 
  & = -\gamma \abs{\omega} \,,
\end{align*}
with the dimensionless strength of the spin fluctuations \cite{ACS03}
\begin{align*}
  \gamma & \equiv 2\pi JN g^2
  = \frac{g^2 N}{2\pi v_F^2\sin\phi_0}
  = \frac{g^2N}{4\pi v_xv_y} \,.
\end{align*}
The inverse spin propagator resummed in the random-phase approximation
has dynamical exponent $z=2$,
\begin{align*}
  \chi^{-1}(i\omega,\q) & = m + \gamma\abs\omega + \nu\abs{\q-\qc}^2 \,,
\end{align*}
where the mass term $m$ measures the distance from the quantum
critical point and $\nu \simeq g^2/\epsilon_F$. Near criticality
$m\approx 0$, the momentum exchanged by scattering off a spin wave is
peaked near $\qc$.  From here on, we shall denote by $\q$ the
deviation from $\qc$.  In the commensurate case there are
logarithmic singularities in the clean bosonic self-energy (from
contracting two spin lines diagonally in Fig.~\ref{fig:b4clean}) that
may lead to an anomalous dimension of the spin propagator
\cite{ACS03}.

\fig{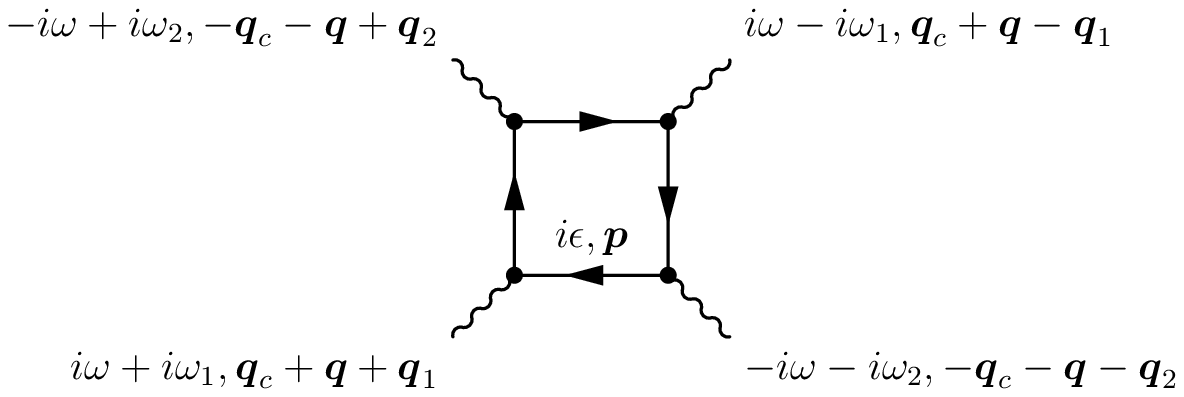}{fig:b4clean}{The clean fermionic four-loop $b_4$,
  indicating the notation for the external frequencies and momenta.}

The 4-point function is given by the fermion loop with four spin
insertions, which provides the bare two-spin interaction:
\begin{align}
  \label{eq:b4clean}
  b_{4} & = -\pi Jg^4 \sum_\alpha  
  \frac{\abs{\omega_1+\omega} + \abs{\omega_1-\omega}
    - \abs{\omega_2+\omega} - \abs{\omega_2-\omega}}
  {[i(\omega_1+\omega_2)-\xi_{\q_1+\q_2}^\alpha]
    \, [i(\omega_1-\omega_2)-\xi_{\q_1-\q_2}^{\bar\alpha}] } \,,
\end{align}
where we have labeled the three independent external frequencies and
momenta as shown in Fig.~\ref{fig:b4clean}.  This is a nonanalytic
function whose value for $\omega\to 0,\q\to 0$ depends on the order of
the limits: it vanishes in the static limit ($\omega\to 0$ first)
while it diverges as $1/\omega$ in the dynamic limit ($\q\to 0$
first).  There is an important difference from the forward-scattering
loop with small external momenta \cite{NM98}: here, symmetrization of
the external lines does not lead to loop cancellation, i.e., a
reduction of the leading singularity.  Instead, symmetrization only
modifies the prefactors but does not change the scaling dimension.

\subsection{Scaling of the clean fermionic loops}

We recall the scaling behavior of the $2n$-point functions
\cite{AC04}.  Since no cancellation of the leading singularity occurs,
we only need to consider the power of external frequency and momentum
and not the particular linear combinations of frequencies and momenta
involved.  Introducing a symbolic notation where $\omega$ denotes a
\emph{positive} linear combination of external frequencies and $q$ a
linear combination of momenta, the $2n$-point functions have the
scaling form
\begin{align*}
  b_{2n} & \sim \frac{g^{2n}}{v_F^2} \,
  \frac{\omega}{(\omega+iv_Fq)^{2(n-1)}} \,,
\end{align*}
where an average over hot spots is understood, which leads to a real
positive function of the frequencies and momenta represented by
$\omega$ and $q$.  For the purpose of scaling, we have substituted $J
\sim 1/v_F^2$ and $\gamma \sim g^2/v_F^2$.

To estimate the relevance of the vertices in the scaling limit
$\omega^{2/z} \sim q^2 \to 0$, where $q$ dominates $\omega$ in the
denominator for $z>1$, consider the $\phi^{2n}$ term in the effective
action \cite{AC04}:
\begin{align*}
  g_{2n} \int (d^2q\,d\omega)^{2n-1} \,
  \frac{\omega}{(v_Fq)^{2(n-1)}} \, \phi^{2n} \,,
\end{align*}
where $g_{2n}$ is the coupling strength related to the vertex function
$b_{2n}$.  Using the scaling dimension of the field
$[\phi^2]=-(d+z+2)$ (in frequency and momentum space) in two dimensions,
\begin{align*}
  [g_{2n}] & = -(2n-1)(2+z) - [z - 2(n-1)] - n(-4-z) = (2-z)n \,.
\end{align*}
The scaling dimension of \emph{all} $2n$-point functions is zero for
$z=2$, i.e., all bosonic vertices are marginal in the scaling limit,
and it is not clear how to perform calculations with such an action.
Our aim is to see if and how the disorder present in real systems
changes the scaling dimension of the fermionic loops.


\section{Disorder corrections to a single fermion loop}
\label{sec:disord}

We consider static impurities modeled by a random local potential with
mean squared amplitude $u^2$ \cite{AGD75}.  As long as no spin-vertex
insertions appear between impurity scatterings, the disorder
corrections in the Born approximation have the standard form.  In the
fermionic propagator $G(i\epsilon,\p) = (i\t\epsilon - \xi_\p)^{-1}$
the Matsubara frequency $\epsilon$ is cut off as $\t\epsilon \equiv
\epsilon + \sgn(\epsilon)/(2\tau)$ at the scale of the impurity
scattering rate $1/\tau = 2\pi \rho_0 u^2$.  Although the
particle-hole bubble $B(i\epsilon+i\omega,i\epsilon,\q)$ with small
momentum transfer $\q$ is cut off by disorder, the direct ladder
resummation has the diffusive form $L(i\epsilon+i\omega,i\epsilon,\q)
\approx u^2/(\abs\omega\tau + Dq^2\tau)$ if both frequencies lie on
different sides of the branch cut on the real line.  The diffusion
constant is $D=v_F^2\tau/2=v_F\ell/2$ in $d=2$.  Throughout this work
we assume that impurity scattering is weak, $1/(k_F\ell) =
1/(\epsilon_F\tau) \ll 1$.

\subsection{2-point function}

Due to the linearized dispersion around the hot spots, the bosonic
self-energy is unchanged by disorder corrections to the fermionic
propagators:
\begin{align*}
  \Sigma_\dirty^{(0)}(i\omega,\qc+\q)
  & = \parbox{80mm}{\includegraphics*{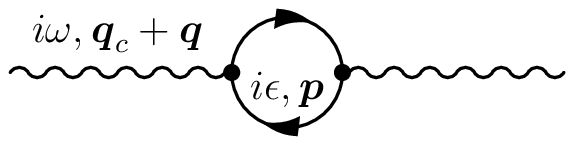}} \\
  & = -\sum_{\text{spin}} g^2 \int \frac{d\epsilon}{2\pi} \,
  \frac{d^2p}{(2\pi)^2} \, 
  \frac{1}{i\t\epsilon - \xi_{\p}} \,
  \frac{1}{i(\wt{\epsilon+\omega}) - \xi_{\p+\qc+\q}} \\
  & = -\gamma \abs\omega \,.
\end{align*}
As the direct impurity ladder with large momentum transfer $\qc$ is
not singular, the leading vertex correction is given by a single
impurity line across the bubble:
\begin{align}
  \Sigma_\dirty^{(1)}(i\omega,\qc+\q)
  & = \parbox{40mm}{\includegraphics*{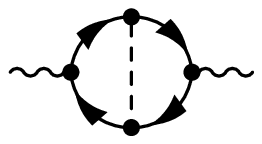}} \notag \\
  & = -2g^2 \int \frac{d\epsilon}{2\pi} \, \frac{d^2p}{(2\pi)^2} \,
  \frac{1}{i\tilde\epsilon-\xi_\p} \,
  \frac{1}{i\tilde\epsilon-\xi_{\p+\qc}} \notag \\
  & \qquad \quad \times u^2 \int \frac{d^2p'}{(2\pi)^2} \, 
  \frac{1}{i\tilde\epsilon-\xi_{\p'}} \,
  \frac{1}{i\tilde\epsilon-\xi_{\p'+\qc}} \notag \\
  \label{eq:masscorr}
  & \approx -\gamma \, \frac{1}{\tau} \;,
\end{align}
where the cutoff scale $\epsilon_F$ is used for the divergent
frequency integral.  The mass correction shifts the position of the
critical point as a function of the control parameter by a finite
amount.  We assume that the system can be fine-tuned again to the
critical point by adjusting the control parameter.

The correction to the bosonic self-energy from the maximally crossed
ladder (cf.\ equation \eqref{eq:Sigmacrossed}) is
\begin{align*}
  \Sigma_\dirty^\cross(i\omega,\qc+\q)
  & \sim \gamma\abs\omega
  \frac{\ln(\abs\omega\tau)}{k_F\ell} \,.
\end{align*}
The logarithm does not change the bare scaling dimension, however the
absence of a corresponding term in $\q$ could tend to increase $z$ and
possibly make the clean vertices irrelevant starting from $z=2$.  We
will not further discuss this possibility and eliminate the crossed
ladders by a small magnetic field because, as we shall see, the
vertices in the presence of impurities will acquire new singularities
due to diffusive ladders which will not be affected by the value of
$z$.

\subsection{4-point function}

Including only self-energy disorder corrections cuts off the
frequencies in the denominator by $\dt\omega = \omega +
\sgn(\omega)/\tau$ and yields the scaling form (where $\omega$ and $q$
represent the same linear combinations of external frequencies and
momenta as in equation \eqref{eq:b4clean})
\begin{align*}
  b_{4,\dirty}^{(0)} & = -\pi Jg^4 \sum_\alpha
  \frac{\omega}{(i\dt\omega-\xi_\q^\alpha)^2}
  \sim \frac{g^4}{v_F^2} \, \frac{\omega}{(\omega+1/\tau+iv_Fq)^2} \;.
\end{align*}
For $\omega,\,v_Fq \ll 1/\tau$, this contribution vanishes linearly in
$\omega$ and is, therefore, irrelevant in the scaling limit.  Instead,
a singular contribution is obtained by including diffusive ladders
into the loop.  Since direct ladders between two propagators with
momenta separated by $\qc$ are not diffusive, we insert direct ladders
only between propagators with almost equal momenta---i.e., near the
same hot spot (see Fig.~\ref{fig:b4dirty}).

\fig{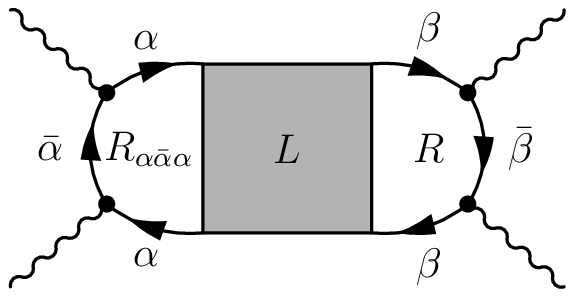}{fig:b4dirty}{The four-point vertex with one direct
  ladder insertion factorizes into pairs of spin vertices with three
  propagators between them ($R$ vertices) and the direct ladder
  ($L$).}

The $R_{\alpha\bar\alpha\alpha}$ subdiagram is constructed from two
propagators near one hot spot $\alpha$ and one propagator near the
associated hot spot $\bar\alpha$, with two spin vertices in between:
\begin{align}
  R_{\alpha\bar\alpha\alpha}(q,q')
  & = g^2 \int \frac{d^2p}{(2\pi)^2} \,
  \frac{1}{i(\wt{\epsilon+\omega})-\xi_{\p+\q}} \,
  \frac{1}{i(\wt{\epsilon+\omega'})-\xi_{\p+\qc+\q'}} \, 
  \frac{1}{i\t\epsilon-\xi_\p} \notag \\
  \label{eq:Raba}
  & = -2\pi i Jg^2 \sum_\alpha
  \frac{\Theta(-\epsilon[\epsilon+\omega])}
  {|\dt{\omega}|+i\xi_\q^\alpha} \, 
  \sgn(\epsilon+\omega') \,,
\end{align}
which saturates to a constant for small $\omega$, $\q$.  Combining the
parts, we obtain for the 4-point vertex with one direct ladder (by the
superscript we denote the number of ladders)
\begin{align}
  b_{4,\dirty}^{(1)} & = -\sum_{\text{spin}}
  \int \frac{d\epsilon}{2\pi} \, R(q_1+q_2,q+q_1)
  \, L(q_1+q_2) \, R(q_1+q_2,q+q_2) \notag \\
  \label{eq:b4dirty1}
  & \sim \frac{g^4}{v_F^2} \, \frac{\omega}{(\dt\omega+iv_Fq)^2} \,
  \frac{1/(k_F\ell)}{\omega\tau+Dq^2\tau}
\end{align}
as the scaling form for typical external $\omega$, $\q$.  We have
checked explicitly that symmetrization of the spin insertions does not
change the leading singularity.  Note that the last factor in the
ladder contribution \eqref{eq:b4dirty1} becomes larger than unity only
for $\omega\tau, Dq^2\tau < 1/(k_F\ell)$, while $b_{4,\dirty}^{(0)}$
is cut off for $\omega\tau, Dq^2\tau < 1$.  This appearance of two
crossover scales is a central observation of our work.  Adding the
contributions with zero and one ladder,
\begin{align*}
  b_{4,\dirty} & \sim \frac{g^4}{v_F^2} \times
  \begin{cases}
    \frac{\omega}{(\omega+iv_Fq)^2}
    & (\omega\tau \text{ and/or } q\ell \gg 1) \\
    \omega\tau^2 \,
    \left[ 1 + \frac{1/(k_F\ell)}{\omega\tau+Dq^2\tau} \right]
    & (\omega\tau \text{ and } q\ell \ll 1) \,.
  \end{cases}
\end{align*}
There are, of course, many more ways to insert ladders into the
4-point vertex.  Since we are interested in the most singular
contribution that determines the scaling, we have already excluded
direct ladders between different hot spots on the basis that they are
not diffusive.  However, one could also add a second ladder in
Fig.~\ref{fig:b4dirty} between the $\bar\alpha$ and $\bar\beta$ lines
(with $\bar\alpha=\bar\beta$).  This does not change the singularity,
but gives a much smaller prefactor $1/(k_F\ell)^2$.  In general, all
such crossings of direct ladders yield internal integrations over the
ladder momenta (as in the case of the cooperon ladder) and give at
most logarithmic corrections but not a higher singularity.
Logarithmic corrections do not change the scaling dimension, and
because of the higher order in $1/(k_F\ell)$, all diagrams with
crossings of direct ladders will be neglected.  Note that the
numerical values of the coefficients of course depend on these
diagrams.  We argue further that the insertion of crossed ladders
leads at most to the same singularity as those with direct ladders, up
to logarithmic terms.\footnote{Alternatively, as discussed above in
  connection with $\Sigma_{\rm dirty}^{\rm crossed}$ one could apply a
  small magnetic field, which does not cut off the singularity in the
  clean case and with direct ladders, but suppresses the contribution
  from crossed ladders.}

In conclusion, this leaves us with a much smaller set of diagrams
displaying the leading singularity: \emph{all possible insertions of
  direct ladders between propagators of the same spin (at the same hot
  spot), which do not cross each other}.

For the 4-point vertex, the ladder diagram in Fig.~\ref{fig:b4dirty}
(with summation over hot spots and symmetrization of external lines
understood) is the only one meeting these conditions and is therefore
sufficient to obtain the leading singularity.  In the scaling limit
$\omega\sim q^2$, the ladder correction has the same scaling behavior
(constant) as the clean vertex $g_4$, hence the 4-point coupling
remains asymptotically marginal even when disorder is included.  This
leads to the question how the higher $2n$-point vertices behave.

\subsection{Higher bosonic vertices}

\fig{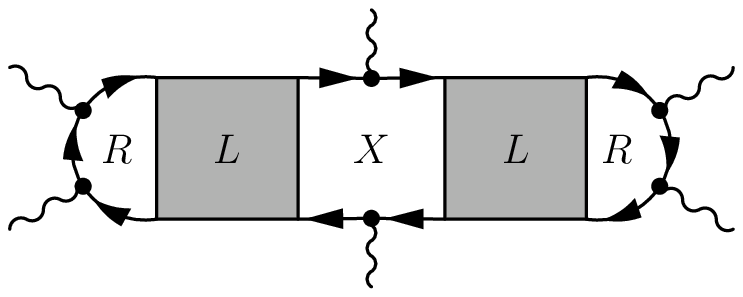}{fig:b2nchain}{The disordered $2n$ loop (shown here
  for $n=3$) with $n-1$ ladder insertions arranged in a chain.  This
  can be extended for larger $n$ by repeating the $(XL)$ block.}

Following the above discussion, we consider a particular insertion of
ladders into the bare $2n$-point vertex which meets the above
condition for a maximally singular contribution.  As shown in
Fig.~\ref{fig:b2nchain}, we group two adjacent spin vertices together
(the $R$ part as in Fig.~\ref{fig:b4dirty}),
\begin{align*}
  R & \sim i\,\frac{g^2}{v_F^2} \, \tau \\
  RL & \sim i\,\frac{g^2}{k_F\ell} \, \frac{1}{\omega+Dq^2} 
  \sim \frac{g^2/\tau}{(\omega+Dq^2)(\epsilon_F)} \;.
\end{align*}
This is connected via a direct ladder $L$ to an $X$ vertex made from
four fermion propagators with two spin and two ladder insertions,
\begin{align*}
  X & \sim \frac{g^2}{v_F^2} \, \tau^2
  \qquad \text{(for small external $\omega$, $\q$)} \\ 
  XL & \sim \frac{g^2}{k_F\ell} \, \frac{\tau}{\omega+Dq^2}
  \sim \frac{g^2}{(\omega+Dq^2)(\epsilon_F)} \;.
\end{align*}
There is another contribution to $X$ not depicted in
Fig.~\ref{fig:b2nchain}, with two spin insertions on the same fermion
line: this term is of the same order of magnitude but generically does
not cancel the one shown.

Repeating the $(XL)$ part $n-2$ times and finishing with another $R$
vertex, we obtain a chain-like diagram with $n-1$ ladders which, for
small $\omega$ and $\q$, can be estimated by the scaling form
\begin{align*}
  b_{2n,\dirty}^{(n-1)}
  & = -\sum_{\text{spin}} \int d\epsilon \, RL(XL)^{n-2}R \\
  & \sim \frac{g^2/\tau}{(\omega+Dq^2)(\epsilon_F)} 
  \left( \frac{g^2}{(\omega+Dq^2)(\epsilon_F)} \right)^{n-2}
  \frac{g^2}{v_F^2}\,\tau\omega \\
  & \sim \frac{g^{2n}}{v_F^2} \, 
  \frac{\omega}{(\omega+Dq^2)^{n-1}(\epsilon_F)^{n-1}} \;.
\end{align*}
Diagrams with fewer than $n-1$ ladders have a weaker singularity and
give an irrelevant contribution to the coupling in the scaling limit.

\fig{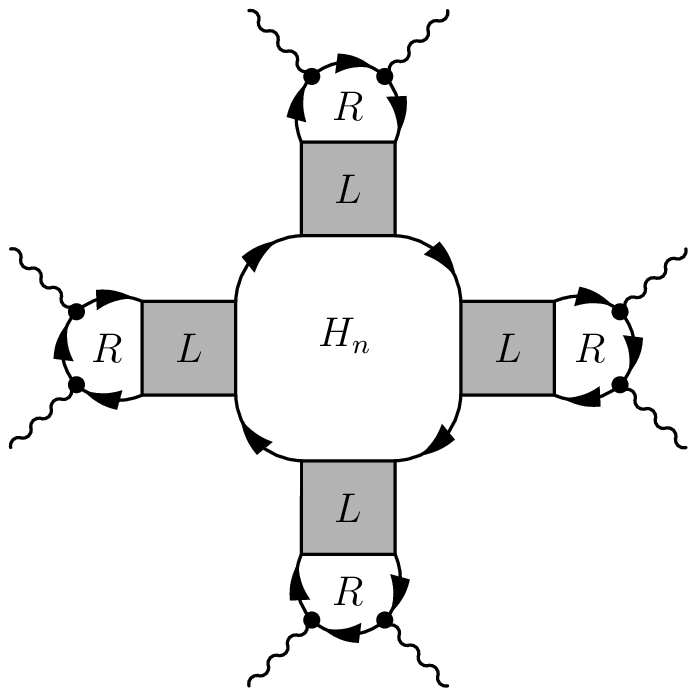}{fig:b2ndirty}{The disordered $2n$ loop for even $n$
  (shown here for $n=4$) with $n$ ladder insertions and a Hikami
  vertex $H_n$ in the middle.}

There are other diagrams with $n$ ladders, which at first appear to be
even more divergent but upon closer inspection turn out to have the
same singularity.  As shown in Fig.~\ref{fig:b2ndirty}, for even $n$
one can connect $n$ $R$ parts via $n$ ladders to an $n$-point Hikami
vertex \cite{Hik81},
\begin{align*}
  H_n & \sim \frac{\epsilon_F}{v_F^2} \, \tau^n \,
  (\omega+Dq^2) \,.
\end{align*}
The additional factor of an inverse diffusion propagator in $H_n$ is
due to the insertion of further single impurity lines which cancel the
constant term and leave only terms linear in $\omega$ and $q^2$ for
small $\omega$ and $q$; this effectively cancels one of the $n$
diffusive ladders.  Connecting the $RL$ parts to the Hikami vertex and
adding the frequency integration along this one large fermion loop,
\begin{align*}
  b_{2n,\dirty}^{(n-1)}
  & = -\sum_{\text{spin}} \int d\epsilon \, H_n (RL)^n
  \sim \frac{g^{2n}}{v_F^2} \,
  \frac{\omega}{(\omega+Dq^2)^{n-1}(\epsilon_F)^{n-1}}
\end{align*}
has the same singularity as the chain-type diagram.

In appendix~\ref{sec:nlsm}, we propose a non-linear $\sigma$ model for
the spin modes coupled to low-energy diffusion modes (instead of the
original electrons) with only one local (constant) coupling.  This
allows us to control all single-loop diagrams exhibiting the leading
singularity, thereby supporting the perturbative calculations in this
work.  Elimination of the diffusive modes would lead again to an
effective action for the spin modes with infinitely many marginal
couplings.


\section{Scaling of the disordered single loops}
\label{sec:scale}

\subsection{Smallness of ladder corrections and second crossover scale}

As for $b_{4,\dirty}$, the disordered loops (beyond the 2-point
function) feature two crossover scales, one where disorder corrections
in the self-energy of the fermion lines cut off the vertices, and
another where ladder corrections lead again to marginal scaling.  The
existence of these two scales can be traced back to the presence of
hot spots.

For comparison, consider the case of forward-scattering bosonic
vertices.  Self-energy disorder corrections become important and cut
off the fermionic propagators at $\omega\tau \approx 1$ and $q\ell
\approx 1$.  Adding one disorder ladder implies adding also two
fermionic propagators with nearby momenta and performing one momentum
integration.  This additional contribution can be estimated as
\begin{align*}
  LG^2 & = 
  \frac{u^2\sqrt{(1+\omega\tau)^2+(q\ell)^2}}
  {\sqrt{(1+\omega\tau)^2+(q\ell)^2}-1} \; 
  \underbrace{\int \frac{d^2p}{(2\pi)^2} \, G^2}_{=2\pi\rho_0\tau}
  = \frac{\sqrt{(1+\omega\tau)^2+(q\ell)^2}}
  {\sqrt{(1+\omega\tau)^2+(q\ell)^2}-1} \; .
\end{align*}
The ladder correction becomes dominant exactly at the same scale
$\omega\tau \approx q\ell \approx 1$ where the self-energy corrections
appear.  Hence, there is only a single crossover scale between clean
and dirty behavior.

Also in the case of backscattering bosonic vertices, the self-energy
corrections set in at $\omega\tau \approx q\ell \approx 1$.  However,
the ladder insertions are modified due to the presence of hot spots
for the two additional fermionic propagators:
\begin{align*}
  LG^2 & = L \;
  \underbrace{JN \int d\xi^\alpha \, d\xi^{\bar\alpha} \, G^2}
  _{=2\pi\rho_0\,\frac{N/(2\sin\phi_0)}{\epsilon_F}}
  = \frac{N/(2\sin\phi_0)}{k_F\ell} \,
  \frac{\sqrt{(1+\omega\tau)^2+(q\ell)^2}}
  {\sqrt{(1+\omega\tau)^2+(q\ell)^2}-1} \; . 
\end{align*}
In comparison with the forward-scattering case above, there is an
additional $\xi$ integration which effectively replaces one factor
$\tau$ by $1/\epsilon_F$, such that the diffusive ladders become
dominant only at a second, lower crossover scale $\omega\tau,(q\ell)^2
\approx 1/(k_F\ell)$.  This means that the impurity ladder scattering
of electron-hole pairs with nearby momenta is less effective by a
factor of $1/(k_F\ell)$ if the particles are forced by spin-vertex
insertions to be near hot spots between impurity ladders.

\subsubsection{Scaling regimes}

According to the above discussion, we can identify three different
scaling regimes for the disordered $2n$-loops:
\begin{align*}
  b_{2n,\dirty} & \approx
  \begin{cases}
    b_{2n,\clean} \sim
    \frac{\omega}{(\omega+iv_Fq)^{2(n-1)}}
    & (\omega\tau,(q\ell)^2 \gg 1) \\
    b_{2n,\dirty}^{(0)} \sim
    \frac{\omega}{(1/\tau^2)^{n-1}}
    & (1 \gg \omega\tau,(q\ell)^2 \gg \frac{1}{k_F\ell}) \\
    b_{2n,\dirty}^{(n-1)} \sim
    \frac{\omega}{(\omega+Dq^2)^{n-1}(\epsilon_F)^{n-1}}
    & (\frac{1}{k_F\ell} \gg \omega\tau,(q\ell)^2)
  \end{cases}
\end{align*}
Note that each pair of fermion-like propagators in the clean case is
replaced by one diffusive propagator and an additional factor of
$\epsilon_F^{-1}$ in the disordered case, leaving $g_{2n}$ marginal in
the scaling limit.  The contributions to the vertices which become
dominant in the different scaling regimes are visualized in
Fig.~\ref{fig:regions} below, and in this figure it is made explicit
how the constraints on $\omega\tau$ and $q\ell$ are to be understood.

There are two different diffusive regimes in the model: a fast one for
charge modes with diffusion constant $D$, and a much slower one for
spin modes with $\nu/\gamma \simeq D/(k_F\ell)$.  We can, therefore,
look at two variants of the scaling limit $\omega \sim q^2$.  For
charge diffusion we have $\omega\simeq Dq^2$ (see the red/dashed
scaling line in Fig.~\ref{fig:regions}), and
\begin{align}
  \label{eq:chargescaling}
  b_{2n,\dirty} & \approx
  \begin{cases}
    b_{2n,\clean}
    \sim \frac{\omega}{(\omega/\tau)^{n-1}}
    & (\omega\tau \gg 1) \\
    b_{2n,\cutoff}
    \sim \frac{\omega}{(1/\tau^2)^{n-1}}
    & (1 \gg \omega\tau \gg \frac{1}{k_F\ell}) \\
    b_{2n,\ladder}
    \sim \frac{\omega}{(\omega\epsilon_F)^{n-1}}
    & (\frac{1}{k_F\ell} \gg \omega\tau)
  \end{cases}
\end{align}
shows a non-monotonous behavior.  On the other hand, if the typical
values of $\omega$ and $\q$ are given by the spin propagators
connected to the external legs of the fermion loops, we put $\gamma
\omega \simeq \nu q^2$ (see the blue/solid scaling line in
Fig.~\ref{fig:regions}), and
\begin{align}
  \label{eq:spinscaling}
  b_{2n,\dirty} & \approx
  \begin{cases}
    b_{2n,\clean}
    \sim \frac{\omega}{(\omega\epsilon_F)^{n-1}}
    & (\omega\tau \gg \frac{1}{k_F\ell}) \\
    b_{2n,\cutoff}
    \sim \frac{\omega}{(1/\tau^2)^{n-1}}
    & (\frac{1}{k_F\ell} \gg \omega\tau \gg \frac{1}{(k_F\ell)^2}) \\
    b_{2n,\ladder}
    \sim \frac{\omega}{(\omega\epsilon_F)^{n-1} (k_F\ell)^{n-1}}
    & (\frac{1}{(k_F\ell)^2} \gg \omega\tau) \;.
  \end{cases}
\end{align}
This scaling analysis suggests that even though the ladder corrections
scale marginally for asymptotically low frequencies, there is a range
of frequencies and momenta where the singular clean vertices are
already cut off and the ladder corrections are still small, such that
the Hertz-Millis theory might apply in this zone.  As we shall see in
the following section, such a regime may be hidden by further
contributions from multiple loops.


\section{Disorder corrections to multiple fermion loops}
\label{sec:multi}

The multi-loop disorder corrections arise from impurity lines
connecting different fermionic loops.  In the simplest case, one takes
$n$ static particle-hole bubbles with two spin insertions each (mass
terms) and connects them with single impurity lines.  As the impurity
lines do not carry frequency, there are only $n$ independent
frequencies in this $2n$-point spin vertex.  The corresponding
coupling in the action has therefore a different scaling dimension
than the single-loop contribution (with $2n-1$ independent
frequencies) and is generally more relevant.  In fact, the missing
frequency integrations lead to a scaling as in the classical field
theory and the Harris criterion \cite{Har74} applied to the bare
model implies that such contributions are relevant in $d<4$.

We first define the $n$-loop vertices $\Delta_{2n}$ in the disordered
spin-fermion model and then discuss the energy scales where these
additional vertices become quantitatively more important than the
single-loop contributions.  Let $\Delta[V]$ denote the particle-hole
bubble at arbitrary momentum transfer $q$ (not just near $\qc$) and
zero external frequency $\omega=0$ in the presence of a particular
configuration of the disorder potential $V$,
\begin{align*}
  \Delta[V] & \equiv \parbox{3cm}{\includegraphics{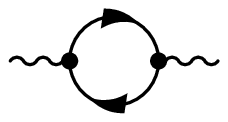}} \\
  & = -\int \frac{d\epsilon}{2\pi} \Tr \left(
    \frac{1}{i\epsilon-\xi-V} \, \Gamma \,
    \frac{1}{i\epsilon-\xi-V} \, \Gamma^* \right)
\end{align*}
where $\xi$ is the hopping matrix, the spin-fermion vertex
$\Gamma=\exp(-iqx)$ is a diagonal matrix in real space (with spin
indices suppressed), and the trace runs over spatial indices.  We
diagonalize $\xi+V = \sum_k \ket{k} \xi_k \bra{k}$ and perform the
$\epsilon$ integration,
\begin{align*}
  \Delta[V] & = -\int \frac{d\epsilon}{2\pi} \sum_{kl}
  \frac{1}{i\epsilon-\xi_k} \bra{k} \Gamma \ket{l} 
  \frac{1}{i\epsilon-\xi_l} \bra{l} \Gamma^* \ket{k} \\
  & = -\sum_{kl} \abs{\bra{k}\Gamma\ket{l}}^2
  \frac{\Theta(-\xi_k\xi_l)}{\abs{\xi_k - \xi_l}} \;.
\end{align*}
We now specialize to the case $\q_i=\qc$ and define the static
connected $2n$-point vertex $\Delta_{2n}$ as
\begin{align*}
  \Delta_{2n}
  & \equiv \left\langle\left(
      \parbox{2.2cm}{\includegraphics{b2small}}
    \right)^n\right\rangle_{\text{disorder average, connected part}}
\end{align*}
We assume that for a generic large momentum transfer near $\qc$ and
generic band dispersion, all $\Delta_{2n}$ have a finite and nonzero
limit as $\omega_i,\abs{\q_i-\qc}\to 0$.  The corresponding terms in
the action
\begin{align*}
  \delta_{2n} \int (d\omega)^n (d^dq)^{2n-1} \, 
  \bigl(\phi(\omega_i) \phi(-\omega_i)\bigr)_{\{q_i\}}^n
\end{align*}
have only $n$ independent frequency integrations but $2n-1$ momentum
integrals.  $\delta_{2n}$ stands for the running coupling with bare
value $\Delta_{2n}$.  The constraint on the frequency integration
leads to a scaling dimension
\begin{align*}
  [\delta_{2n}] & = -nz - (2n-1)d - n(-d-z-2) \\
  & = d - n(d-2) && (z=2) \\
  & = 2 && (d=2) \;.
\end{align*}
Thus, infinitely many couplings $\delta_{2n}$ are all equally
relevant.  The singularity found in two dimensions is so strong that
one expects that even in three dimensions $\Delta_4$ remains a
relevant perturbation and destabilizes a Gaussian fixed point in the
disordered model, which may be related to the difficulty encountered
in reconciling experiments with the Hertz-Millis theory for $d=3$,
$z=2$ \cite{Col02,HKV07}.

\fig[height=2.5cm]{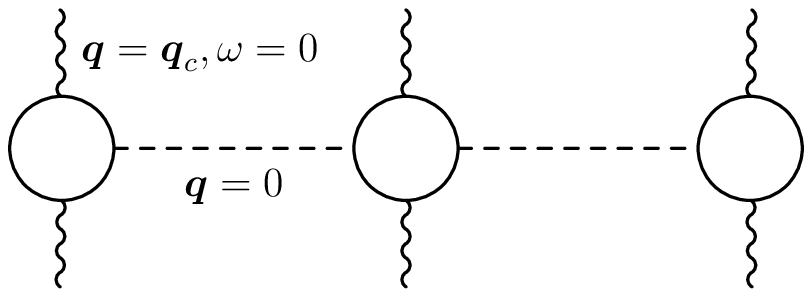}{fig:d2nchain}{Chain-type
  contribution to the $\Delta_{2n}$ vertex (shown here for $n=3$) with
  zero momentum transfer on the impurity lines.}

In order to estimate the quantitative importance of the $\Delta_{2n}$
vertices with respect to the $b_{2n}$ vertices, one needs to find the
contributions to $\Delta_{2n}$ at lowest order in $1/(k_F\ell)$.  For
$n=1$, the leading term is the mass correction in equation
\eqref{eq:masscorr}.  For $n\geq 2$, leading contributions to
$\Delta_{2n}$ are given by chains of $n$ $\Delta[V]$'s connected by
$n-1$ single impurity lines with zero momentum transfer (see
Fig.~\ref{fig:d2nchain}).  The terminal bubbles of the chain (with one
impurity line) are roughly $g^2/v_F^2$, while the intermediate bubbles
with two impurity lines are approximately $g^2/(v_F^2\epsilon_F)$.
The complete chain can therefore be estimated as
\begin{align*}
  \Delta_{2n} & \simeq \frac{g^2}{v_F^2}\, u^2 
  \left[ \frac{g^2}{v_F^2\epsilon_F} \,
    u^2 \right]^{n-2} \frac{g^2}{v_F^2} \\
  & \simeq \frac{g^{2n}}{v_F^2} \, \frac{1}{\epsilon_F^{n-2}} \,
  \frac{1}{(k_F\ell)^{n-1}} \quad (n\geq 2) \;.
\end{align*}
In the scaling limit dominated by spin diffusion, we can replace
$(q/k_F)^2$ by $\omega/\epsilon_F$, and the single-loop vertex
functions $b_{2n}$ scale as given in equation \eqref{eq:spinscaling}.
Below a certain frequency scale $\omega$, the relevant couplings
$\delta_{2n}$ will necessarily become larger than the marginal
couplings $g_{2n}$; in order to find this scale one has to compare
$b_{2n,\clean}$ with $\Delta_{2n}/\omega^{n-1}$, to account for the
missing $n-1$ frequency integrations in the couplings $\delta_{2n}$
with respect to $g_{2n}$:
\begin{align*}
  b_{2n,\clean} & \approx
  \frac{\Delta_{2n}}{\omega^{n-1}} \\
  \frac{\omega}{(\omega\epsilon_F)^{n-1}}
  & \approx \frac{1}{\omega^{n-1}\epsilon_F^{n-2}(k_F\ell)^{n-1}} \\
  \omega\tau & \approx \frac{1}{(k_F\ell)^{n-2}} \;.
\end{align*}
Hence, in the spin scaling limit the higher vertices $\Delta_{2n}$
start to dominate the clean single-loop vertices $b_{2n,\clean}$ at
successively lower frequency scales.  Likewise, the cutoff vertex
$b_{2n,\cutoff}$ is of the same magnitude as
$\Delta_{2n}/\omega^{n-1}$ for $\omega\tau \approx
1/(k_F\ell)^{(2n-3)/n}$.  Extending these arguments beyond the
scaling limit to the $\omega$-$q^2$ plane, we obtain the crossover
lines between single- and multi-loop contributions indicated in
Fig.~\ref{fig:regions}. In summary, $\Delta_4$ dominates the clean
vertex $b_4$ below $\omega \lesssim v_Fq/\sqrt{k_F\ell}$ for $v_Fq
\gtrsim 1/\tau$, and the cutoff vertex $b_4$ below $\omega \lesssim
1/(\tau\sqrt{k_F\ell})$ for $v_Fq \lesssim 1/\tau$.  The higher
vertices $\Delta_{2n>4}$ become dominant for $v_Fq \lesssim 1/\tau$ only
below $\omega \approx 1/(\tau k_F\ell)$.

\figs[height=5cm,clip]{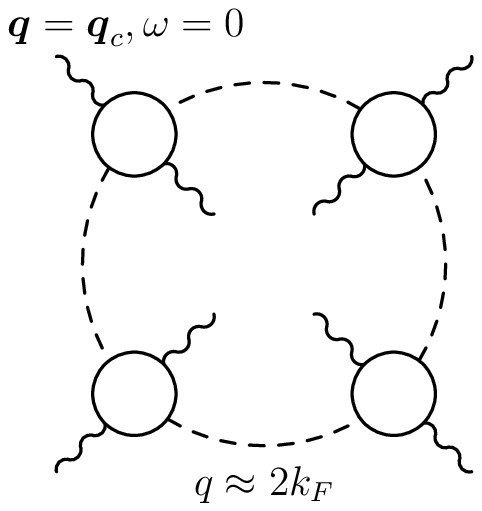}{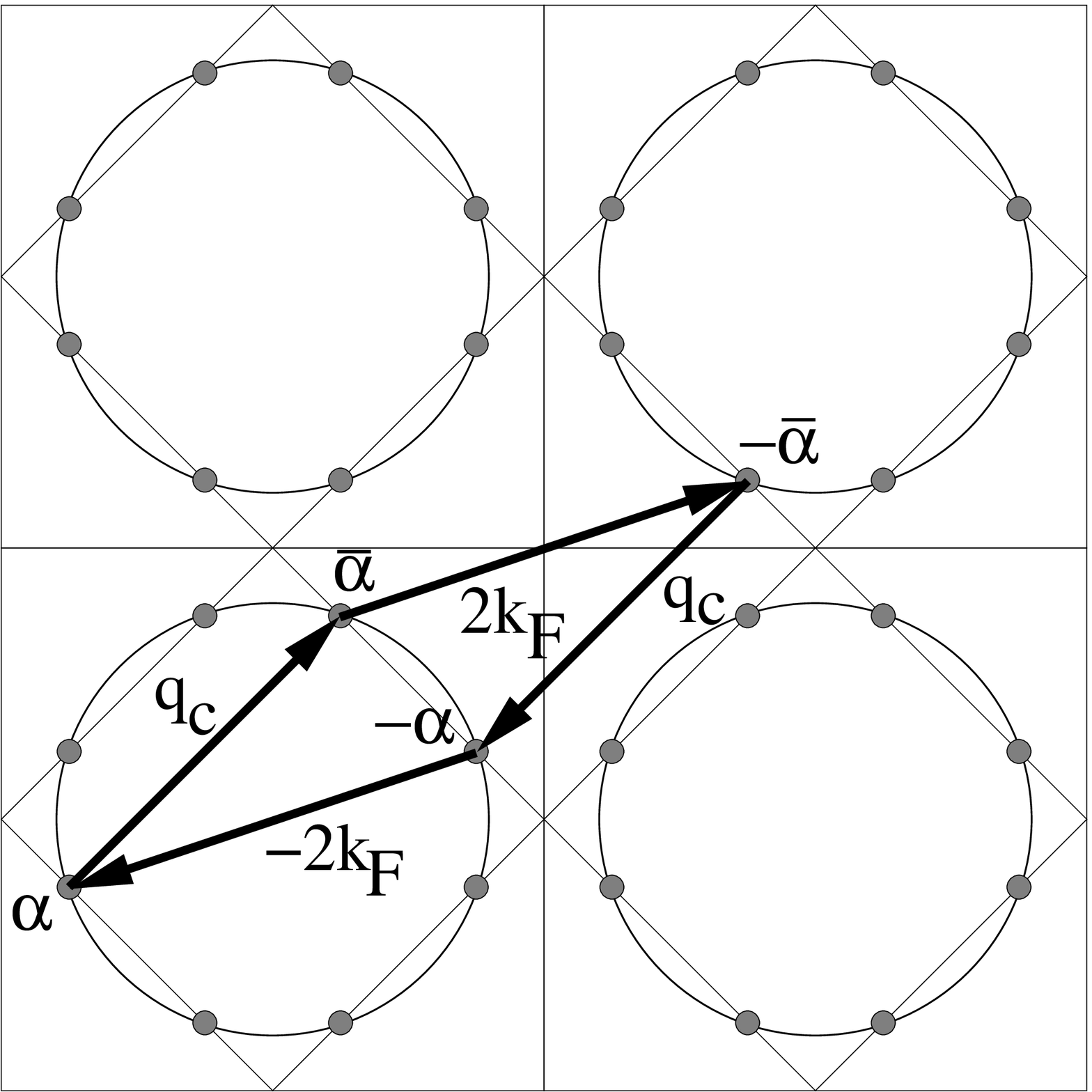}{fig:fs_2kf}{\textbf{(a)}
  Ring-type contribution to the $\Delta_{2n}$ vertex (shown here for
  $n=4$) which is peaked when the impurity lines carry a momentum near
  $2k_F$.  \textbf{(b)} Points on Fermi surface connected in the
  singular $\qc$-$2k_F$ bubbles, with a commensurate $\qc$ and
  incommensurate $2k_F$.}

In addition to the chain-type diagrams above, there are diagrams with
$n$ bubbles arranged in a ring and connected by single impurity lines.
In this case, one has to integrate over the momentum carried by the
impurity lines, involving also momenta near $2k_F$ where the bubble
with two static $\qc$ spin insertions and two static $2k_F$ charge
insertions becomes singular in the clean case (see
Fig.~\ref{fig:fs_2kf}).  This singularity is related to the well-known
$2k_F$ singularity of the particle-hole bubble \cite{AIM95}; while in
the clean case only a small region of the Fermi surface around the hot
spots is visited, impurity scattering visits the whole Fermi surface,
including the parts separated by $2k_F$.  The disorder correction to
the fermionic self-energy provides a cutoff for this singularity, thus
changing the estimated power of $1/(k_F\ell)$ in the expression for
$\Delta_{2n}$.  We estimate that all such diagrams are at least of
order $1/(k_F\ell)^2$, which implies that they become dominant at the
same scale as $\Delta_6$, i.e., only below $\omega \approx 1/(\tau
k_F\ell)$.

Therefore, in the spin scaling limit we still obtain clean anomalous
behavior above $\omega \simeq 1/\tau$, which gets modified by a single
relevant vertex $\Delta_4$ for $1/\tau > \omega > 1/(\tau k_F\ell)$.
For $\omega < 1/(\tau k_F\ell)$, ever higher multi-loop vertices
$\Delta_{2n}$ add to the single-loop vertices.

\fig[height=10cm,clip]{regions}{fig:regions}{(Color online)
  Regions in the $\omega$-$q^2$ plane where different contributions to
  the spin vertices become dominant.  The lines separating the
  different regions are meant as a guide only, as they are determined
  only up to prefactors of order $\O(1)$.}

The problem of a disordered AFM was considered in Ref.~\cite{KB96},
where the authors analyzed an AFM $\phi^4$ model with a local
$u\phi^4$ interaction and a random mass term $m\phi^2$ within an
$\epsilon=4-d$ expansion.  Averaging over random mass configurations
yields a $\Delta(\phi^2)_\omega(\phi^2)_{\omega'}$ term from the
static mass-mass correlation $\vev{mm}$, which depends on three
independent momenta but only two frequencies.  Therefore, $\Delta$ is
a relevant coupling in $d<4$ and the clean model is unstable against
arbitrarily small disorder.  Recently, this model has also been
studied using a strong-disorder RG \cite{HKV07}.

However, according to the above discussion the clean AFM $\phi^4$
model is not applicable in $d=2$ because $u$ is nonlocal and there are
infinitely many additional marginal couplings which even change the
direction of the RG flow for $u$ \cite{AC04}.  Furthermore, we have
pointed out that for a generic fermionic band dispersion the average
over a random fermionic potential implies a fluctuating mass term
which has non-vanishing higher cumulants $\vev{m^n}$.  While these
higher cumulants are irrelevant in $d=4$, in two dimensions all
couplings $\Delta_{2n}$ are equally relevant and may possibly change
the direction of the RG flow also for $\Delta=\Delta_4$.  Thus, it is
not obvious which of the results of \cite{KB96} hold in $d=2$.


\section{Discussion and conclusions}
\label{sec:discus}

We have shown that the fermionic loops with large momentum transfer,
which are relevant to describe an AFM transition, exhibit two
crossover energy scales if disorder corrections are added.  In the
scaling limit set by charge diffusion $\omega\simeq Dq^2$, the
singularities of the clean (marginal) vertices $g_{2n}$ are cut off at
scale $\omega \approx 1/\tau$ and the 4-point vertex and beyond become
irrelevant.  Below $\omega \approx 1/(\tau k_F\ell)$, however,
diffusive ladders lead again to marginal scaling, albeit with a
diffusive functional form of the vertex functions different from the
clean case.  On the other hand, in the scaling limit set by the slower
spin diffusion $\omega\simeq \frac{\nu}{\gamma} q^2 \simeq
Dq^2/(k_F\ell)$, these crossovers occur at the same momentum scales
but at frequency scales smaller by a factor of $1/(k_F\ell)$.  A
summary of crossover scales and boundaries is reported in
Fig.~\ref{fig:regions}.

In addition to the single fermion loops, there are diagrams made of
multiple fermion loops connected only by impurity lines, which were
not present in the clean case.  In accordance with the Harris
criterion applied to the bare model, these are relevant perturbations
in $d=2$ which make the clean model unstable against disorder.
Indeed, in contrast to $d=4$, in two dimensions the bosonic model
contains infinitely many equally relevant vertices $\Delta_{2n}$ due
to disorder corrections to multiple fermion loops, which we estimate
to become important below $\omega \simeq 1/(\tau\sqrt{k_F\ell})$
(charge diffusion) or $\omega \simeq 1/\tau$ (spin diffusion),
respectively.

Combining these results, we can distinguish two cases: if scaling is
dominated by charge diffusion, the anomalous clean behavior
\cite{ACS03} above $\omega \approx 1/\tau$ is cut off below to make
place for an essentially non-interacting behavior (Hertz-Millis theory
with only irrelevant couplings) until $\omega \approx
1/(\tau\sqrt{k_F\ell})$, where the relevant disorder vertex $\Delta_4$
starts to dominate.  Below $\omega\approx 1/(\tau k_F\ell)$,
infinitely many more disorder vertices $\Delta_{2n}$ dominate the
respective single-loop vertices.

In the case of scaling determined by spin diffusion, the clean
anomalous behavior is modified below $\omega \approx 1/\tau$ by a
single disorder vertex $\Delta_4$, while the higher single-loop
vertices $g_{2n>4}$ are not yet strongly modified by disorder.  Only
below $\omega \approx 1/(\tau k_F\ell)$ the single loops are cut off,
but at the same time more disorder vertices $\Delta_{2n>4}$ appear.
In consequence, there is a direct crossover from clean anomalous to
strongly disordered behavior which implies that the theory of Hertz
and Millis may not be applicable for the two-dimensional
antiferromagnet.

TE wishes to thank A. Chubukov, A. Rosch and M. Salmhofer for fruitful
discussions.  We thank the Alexander von Humboldt foundation (TE and
CDC), and the Italian Ministero dell'Universit\`a e della Ricerca
(PRIN 2005, prot.\ 2005022492) for financial support.


\appendix
\section{Non-linear sigma model}
\label{sec:nlsm}

In this appendix we outline the steps to formulate the problem of
disordered single loops in terms of the non-linear $\sigma$ model for
interacting disordered electrons \cite{Weg79,ELK80,Fin83,BK94} in
order to support the perturbative calculations in this work and
control all contributions with the leading singularity.  Assuming some
familiarity with the non-linear $\sigma$ model itself, we introduce
only the modifications necessary to accommodate the spin-vertex
insertions.  We start with noninteracting electrons in the presence of
disorder, add the spin vertices as couplings to an external field,
perform the disorder average using the replica method, integrate over
the fermionic degrees of freedom and obtain an effective action for
the $Q$ matrices (in standard notation),
\begin{align*}
  S[Q,\phi] & \simeq \int dr
  \left\{ \tfrac{\pi\rho_0}{8\tau} \Tr Q^2
    - \tfrac 12 \Tr \ln
    \bigl( G_0^{-1}+\tfrac{i}{2\tau}Q-\phi \bigr) \right\}
\end{align*}
where $Q$ are matrices in frequency and replica space with a weak
real-space dependence representing the electron-hole pairs while
$\phi(r)=\t\phi(r) \exp(ir\qc)$ is the staggered external field, with
the matrix $\t\phi$ slowly varying in space.\footnote{For simplicity,
  we do not write explicitly the frequency dependence of $\t\phi$.}
The $Q$ matrices can be expressed as a rotation $Q = T^{-1}Q_\sp T$ of
the saddle-point solution $Q_\sp = \sgn(\epsilon)$ of the classical
action for $\phi=0$.  In the vicinity of the saddle point, one obtains
\begin{align*}
  S[Q,\t\phi]
  & \simeq \int dr
  \bigl\{ D \Tr (\nabla Q)^2 - 4 \Tr(\epsilon Q)
  - \text{(terms in $\t\phi$ and $Q$)} \bigr\} \;,
\end{align*}
where the first two terms are the standard non-linear $\sigma$ model
for disordered electrons and additional terms are obtained by
expanding the logarithm in $\t\phi$.  One thus obtains vertices
$(Q\t\phi)^k$ which couple the diffusons to the external spin field.

We parametrize the $Q$ matrices as $Q = e^{W/2} Q_\sp e^{-W/2}$, where
the diffuson propagator $\langle WW \rangle $ is represented by the
direct ladder $L$.  There is a term $Q_\sp \t\phi W \t\phi$ in the
action which corresponds to the $R$ vertex in equation
\eqref{eq:Raba}, and terms $W \t\phi W \t\phi$ and $W^2 \t\phi^2$
(corresponding to the $X$ vertex) with two diffusons connected to two
spin-vertex insertions.  Higher vertices $(Q\t\phi)^k$ with more than
two spin insertions generate sub-leading contributions by the scaling
arguments presented below.

For the leading singularity it suffices to consider all possible ways
to connect $X$ and $R$ vertices via diffusons, with the possible
inclusion of the Hikami vertices $H_k$ \cite{Hik81}, which represent
the interaction of the diffusons in the absence of $\t\phi$.  Each
vertex contributes a $\delta$ function of all momenta, while each
ladder implies an integration over its momentum.  Hence, the power
counting depends only on the number $n_\delta$ of additional $\delta$
functions beyond the overall $\delta$ function of external momenta,
which is the number of vertices minus one, $n_\delta = n_V - 1 = n_R +
n_X - 1$.  The Hikami vertices do not contribute to $n_\delta$ since
they scale as an inverse ladder.  The number $2n$ of spin insertions
determines $n_R + n_X = n$, such that $[g_{2n}] = z - 2n_\delta = z -
2(n-1)$ in accordance with our previous calculation.  This maximal
scaling dimension is valid for a large class of diagrams, two
representatives of which are the chain-type diagram in
Fig.~\ref{fig:b2nchain} and the star-shaped diagram in
Fig.~\ref{fig:b2ndirty}.

The class of diagrams with leading singularity contains, however,
contributions with a different relative importance measured in powers
of $1/(k_F\ell)$.  The dominant terms of $\mathcal{O}
(1/(k_F\ell)^{n-1})$ correspond to a single fermion line and no
crossings of direct ladders, otherwise an additional factor
$1/(k_F\ell)^l$ is generated according to the formula
\begin{align*}
  l & = \sum_{k>1} n_{H_{2k}} (k-1) - \frac{n_R}{2} + 1
\end{align*}
which is valid for any \emph{connected} diagram with $n_{H_{2k}}\geq
0$ Hikami vertices $H_{2k}$ and $0\leq n_R\leq n$ vertices $R$ as well
as $n_X = n-n_R$ vertices $X$.  For instance, Fig.~\ref{fig:b4dirty}
corresponds to $RLR$ and has $l=0$ (with $n_{H_{2k}}=0$ and $n_R=2$),
while an additional ladder crossing the first one corresponds to
$\Tr(XLXL)$ with a single closed fermion loop which has $l=1$ (with
$n_R=0$).

If we include the spin propagator in the action and integrate over the
diffusons, the purely spin-wave action is recovered with its infinite
number of marginal $\t\phi$ vertices, while integrating over $\t\phi$
will likely lead to an action with an infinite number of marginal
diffuson vertices.  It appears that the action written in terms of
both $\t\phi$ and diffusons provides the simplest formulation of the
interaction of the low-energy spin and diffusive modes, with only one
(constant) coupling associated to the coupling term $\int dr \Tr
[Q\t\phi Q\t\phi]$.

\section{Crossed impurity diagrams}
\label{sec:crossing}

In this appendix we consider the disorder corrections to the bosonic
self-energy beyond the Born approximation due to maximally crossed
ladders $L_c$.  In contrast to the direct ladders they can have a
diffusive contribution also between two propagators separated by an
incommensurate $\qc$ (e.g., in the bosonic self-energy), and as the
ladder momentum is integrated over, this typically results in a
logarithm:
\begin{align}
  \Sigma_\cross(i\omega,\qc+\q) & = -2g^2 \int \frac{d\epsilon}{2\pi}
  \int \frac{d^2p}{(2\pi)^2} \int \frac{d^2p'}{(2\pi)^2} \,
  G_{\p+\q}^{\alpha}(\epsilon+\omega) \, 
  G_{\p}^{\bar\alpha}(\epsilon) \notag \\
  & \qquad \times L_c(\omega,\p+\p') \,
  G_{\p'+\q}^{-\bar\alpha}(\epsilon+\omega) \, 
  G_{\p'}^{-\alpha}(\epsilon) \notag \\  
  \label{eq:Sigmacrossed}
  & \overset{(\omega\tau \ll 1)}{\longrightarrow}
  -\frac{g^2}{v_F^2} \, \frac{1}{k_F\ell} \, \abs\omega 
  \left( \frac{4}{N\pi} \,
    \frac{-\ln(\abs{\omega}\tau)}{1+\t Dq^2\tau} \right) \,.
\end{align}
While the above expression $\propto \abs\omega\ln(\abs\omega\tau)$
vanishes for $\abs\omega\to 0$, it is logarithmically larger than the
Landau damping term $\propto \abs{\omega}$ and has the same sign.
Thus, $\Sigma_\cross$ enhances the frequency-dependent part of the
bosonic propagator while leaving the momentum-dependent part unchanged
for small frequencies and momenta.  Therefore, there may be a tendency
to increase $z$ beyond $z=2$.  On the other hand, for $N=8$ the term
in parentheses becomes of $\O(1)$, i.e., comparable with the direct
ladder contribution, only if $\abs\omega \lesssim 10^{-3}\tau$.  As
previously mentioned, in this work we discarded these contributions
assuming that a small magnetic field would cut off this logarithmic
singularity.


\end{document}